\documentclass[10pt]{article}
\usepackage[a4paper]{geometry}
\usepackage[T1]{fontenc}
\usepackage[utf8x]{inputenc} 
\usepackage[affil-it]{authblk}
\usepackage{lmodern}
\usepackage[english]{babel}
\usepackage{amsfonts}
\usepackage{amssymb}
\usepackage{ stmaryrd }
\usepackage{ textcomp }
\usepackage{ bbold }
\usepackage{amsmath}
\usepackage{authblk}
\usepackage[toc,page]{appendix}
\usepackage{bbm} 
\usepackage{hyperref}
\usepackage{ mathrsfs }
\usepackage[all]{xy} 
\usepackage{xcolor}
\usepackage{tikz} 
\usepackage{babel}
\usepackage{amsthm}
\usepackage{verbatim}
\usepackage{listings}
\PrerenderUnicode{é}
\usepackage{amsmath}
\newcounter{eqn}

\makeatletter
\newcommand{\putindeepbox}[2][0.7\baselineskip]{{%
    \setbox0=\hbox{#2}%
    \setbox0=\vbox{\noindent\hsize=\wd0\unhbox0}
    \@tempdima=\dp0
    \advance\@tempdima by \ht0
    \advance\@tempdima by -#1\relax
    \dp0=\@tempdima
    \ht0=#1\relax
    \box0
}}
\makeatother

\date{January 14th 2016}

\title{Essay on the State of Research and Innovation in France and the European Union}
\author{Antoine Kornprobst  \thanks{Electronic address: \texttt{Antoine.Kornprobst@malix.univ-paris1.fr}}}
\affil{Université Paris 1 Panthéon Sorbonne\\ Labex ReFi \footnote{This work was achieved through the Laboratory of Excellence on Financial Regulation (Labex ReFi) supported by PRES heSam under the reference ANR­10­LABX­0095. It benefited from a French government support managed by the National Research Agency (ANR) within the project Investissements d'Avenir Paris Nouveaux Mondes (Investments for the Future Paris ­New Worlds) under the reference ANR­11­IDEX­0006­02.}}

\begin{document}
\maketitle
\tableofcontents
\bigskip

The author gratefully acknowledges the help of \textbf{Maximilien Nayaradou} (Finance Innovation), \textbf{Henry Marty-Gauquié} (Banque Européenne d'Investissement), \textbf{Philippe Roy} (Cap Digital) and \textbf{Emmanuel Kornprobst} (Rouen University of Law) whose guidance while exploring this subject was very much appreciated.

\section{Introduction}

Innovation in the economy is an important engine of growth and no economy, whatever its complexity and degree of advancement, whether it is based on industry, agriculture, high tech or the providing of services, can be truly healthy without innovating actors within it. The aim of this work, done by an applied mathematician working in finance, not by an economist or a lawyer, isn't to provide an exhaustive view of the all the mechanisms in France and in Europe that aim at fostering innovation in the economy and to offer solutions for removing all the roadblocks that still hinder innovation; indeed such a study would go far beyond the scope of this study. What I modestly attempted to achieve in this study was firstly to draw a panorama of what is working and what needs to perfected as far as innovation is concerned in France and Europe, then secondly to offer some solutions and personal thoughts to boost innovation.\\

Those solutions will mostly be articulated through the development of business and research clusters like Finance Innovation or Cap Digital that provide a favorable setting for start-up companies in a particular sector of activity to be born and prosper as well as through some European Institutions like the \textit{European Investment Bank} (EIB) that, among its other missions, helps young companies, especially innovating ones, to obtain all the funding they need to secure their future and to ensure not only the continued existence but a strong expansion of the number of jobs they represent. Innovation, whatever the field of activity, isn't the exclusive domain of start-ups and \textit{Small and Medium Enterprises} (SME) of course, indeed it could be argued on the contrary that a majority of the economic innovation is made by either large multinational companies or the Sate behind large public companies or Universities. Still, it is often in reality through partnerships between large institutional agents and small young companies, or the sale of a start-up to a large company, or the funding by a large company of a researcher's project, that real innovation is made and our study will aim at demonstrating that fact.\\

In France, as in all countries, the state of innovation is best described by a \textit{National System of Innovation} (NSI) that takes into account the actors of innovation (fundamental and applied researchers, innovating entrepreneurs, business and research clusters), the structures of innovation (SME's, multinationals, universities, the Sate, European Institutions, etc.) and the results of innovation (economic growth, lower unemployment).\\

Innovation is always based on research in all its forms (fundamental, experimental, applied...) that lead to the creation of intellectual property. That intellectual property can for example take the form of patents that are submitted in France to the \textit{Institut National de la Propriété Industrielle} (INPI) and which are protected by the law for 20 years, the form of patented inventions (that do not have the same legal status as patents but enjoy equivalent protections) or the form of a know-how protected by the laws governing trade secrets. Then innovation in the economy could be defined as exploiting this intellectual property originating from research to produce goods and services. Actual production is being ensured by public, private or hybrid public/private companies that can either be the original proprietor of the intellectual property by having funded the research themselves, or that can have rented that intellectual property through licensing, or that could have acquired that intellectual property and associated technology by purchasing a small start-up company for instance : Google bought Youtube, Facebook bought What's App to give a few well known examples.\\

An NSI is constituted of six components : Human Resources, Public Research, Private Research, Relationship between Industry and Science, Innovating Entrepreneurs and General State Policy. Each of those components have to be evaluated individually and in relation to one another in order to obtain an overview of the state of innovation in a country. Globally, although France possesses excellent research and a remarkable dynamism of its actors of innovation, its NSI is quite poor according to a 2011 report by the European Commission [3] : it ranks eleventh among the European Union's 27 members at the time (Croatia joined in 2013) and it is classified as an "innovation follower" rather than an "innovation leader". France possesses excellent NSI components which tend unfortunately to fail at interacting with each others and that results in a lot of red tape that stifles innovation. \\

\section{The Tools of Innovation and Economic Growth}
\subsection{The Legal Framework of Innovation}
\subsubsection{Public Research and Innovation Structures}

The first  important characteristic of research in France that differs from almost all of the other OECD member states is the key part of the State in its financing. Indeed, the part of the French Sate is around 50 \% of all research and development expenses in France, compared to around only 30 \% in Germany and the U.K, according to a 2014 report by the OECD [2]. Moreover, those research expenses aren't limited to traditionally regal domains like national defense or nuclear energy but extends to civilian research and development. The second exclusive characteristic of French research is the relatively minor role played in public research by the universities themselves : they yield the most important role to large public research administrations which are recognized to be among the best in the world in their respective fields of expertise. Those public research administrations are for example the CNRS (\textit{Centre National de la Recherche Scientifique / National Center for Scientific Research}), the CEA (\textit{Centre d'Études Atomiques / Center for Nuclear Research}), the CNES (\textit{Centre National d'Études Spatiales / National Center for Space Research}), the INRIA (\textit{Institut National de Recherche en Informatique et en Automatique / National Institute for Computer Science}), the INSERM \textit{(Institut National de la Santé et de la Recherche Médicale / National Institute for Health and Medical Research}), the CNET (\textit{Centre National d'Études des Télécommunications / National Center for Information Technology}), and many others so that each major field of public research has its own public administration.\\

The CNRS for example has over 33,000 people on its payroll and administrates over 1,100 research laboratories (including 40 located abroad). Its annual budget is of 3.3 billion euros including almost 700 million euros of its own resources which means money that comes from contracts with private sector counter-parties, public institutions like the ANR (Agence Nationale de la Recherche / National Agency for Research) or the European Institutions. Between 2007 and 2011, an average of 43,000 peer reviewed publications were produced each year under the CNRS brand and the CNRS also filled between 600 and 800 patents each year. Those patents are often licensed to the private industrial sector that can translate excellence in scientific research into real economic growth. Indeed, the CNRS has a long tradition of establishing joint ventures (3,000 per year on average) with large French industrial actors, public universities, other public research administrations or other private research institutions. The CNRS is also very active as a real economy actor of its own, applying its research and producing economic growth through the creation, under the CNRS brand, of many start-up companies by member of its staff (more than 1,100 start-ups created since 1999).\\

This importance of the State in French research and innovation has traditionally created a "vertical" structure where public research administrations were often associated with either large companies, sometimes partly State owned, of the aeronautics, space, nuclear, transportation, public utilities or telecommunications sector, or on the other hand with start-up companies born of a patent filled by a public research administration. Medium sized companies in the middle were typically excluded of these arrangements, leaving them with only a marginal role to play in research and innovation in the French economy. Moreover, this vertical structure created a tendency to concentrate the top decision and policy making at the ministerial level (Defense Ministry, Transport Ministry, Education Ministry, etc...) which fostered short term political interference to the detriment of a sound business strategy. Indeed, this vertical structure of research and innovation in the French economy wasn't immune to criticism as the 2012  \textit{Rapport de la Mission d'Évaluation Relative au Soutien à l'Économie Numérique et à l'Innovation} (Report of the Committee for the Development of the Economy and Innovation in the Digital Age) [1], published by the Finance Ministry, underlined. Some of its main conclusions were : 
\begin{itemize}
\item There is a vertical compartmentalization of research that, to this day, negatively impacts the training, either initial or during their whole career, of engineers and researchers, thus limiting the potential for interdisciplinary collaborative work, which is essential for true innovation to take place.
\item Since there is a partial disconnection between the world of research, dominated as we have seen by public money and thus geared toward reaching the goals of research and innovation of the public sector, and the world of business and industrial development, the question of the profitability of the main orientations, decided at the top political level, of the French research and innovation policies needs to be raised more often. Only a research and innovation policy that aims at being profitable in the medium to long term has a chance to foster real economic growth in the country.
\item There is to this day too much opacity and too much rigidity in the way research projects are financed and not enough freedom allowed for researchers, especially those who intent on creating start-up companies, to make their project evolve. As a matter of fact, financing is very often tied to a series of rules and criteria established by the State that do not always take into account the evolution of the economy, consumer habits, social structure or the environment.
\end{itemize}

These structural flaws of the traditional research and innovation structure in France are also compounded with the worsening flaws of the French education and higher learning systems and the public universities are often in competition with the \textit{Grandes Écoles} (public or private) which have much more selective entry exams. Public universities, due to lack of material means and questionable policy decisions over the last few decades, tend have sometimes difficulties providing a good level of education to the students, especially at the bachelor level (i.e an education that will enable students to find a job). While the French education system is good at producing world-class researchers, engineers and business people, it lags behind in offering young people proper technical training in the IUT (\textit{Institut Universitaire de Technologie} / Technical Training Universities) and through the deliverance of diplomas such as the BTS (\textit{Brevet de Technicien Supérieur} / Qualified Technician Certificate) which are essential for innovation to work in the real economy. As a matter of fact, that cutting edge start-up company created by a CNRS researcher will need qualified IT technicians, marketing and communication people, etc, to live and develop itself and that's very often what is difficult to find in France, despite persistent high unemployment.\\

Despite all those structural flaws in its research and innovation system, France has been trying since the late 1990' to correct the past mistakes and bring its NSI closer in terms of efficiency to those of its OECD peer countries. Many legislative and administrative decisions were taken to boost the efficiency of French research and innovation and that tendency is accelerating since the advent of the Euro and the ever closer European integration. Among the top measures taken, we could underline :

\begin{itemize}
\item Incentives to increase technological transfers between the public and private sectors through a series of legislative measures since 1994.
\item Creation since 2004 of business and research clusters on the American "Silicon Valley" model.
\item Creation of a very attractive special tax status for the researchers and employees who come and establish themselves in France at the invitation of a French company (\textit{Code Général des Impôts} (CGI) / Tax Code, art.81.B).
\item Creation in 2006 of the ANR and of the AERES (\textit{Agence d'Évaluation de le Recherche et de l'Enseignement Supérieur} / National Agency for the Evaluation of Research and Higher Learning Education) that has since been renamed as  the\textit{ Haut Conseil de l'Évaluation de le Recherche et de l'Enseignement Supérieur} / High Council for the Evaluation of Research and Higher Learning Education.
\item Establishment since 2010 of successive PIA's (\textit{Programmes d'Investissement d'Avenir }/ Programs for Future Investments).
\item Introduction in 2008 of a new legal status for French public universities that grant them a higher degree of autonomy. In particular, the universities are opening up to the public sector with the goal of developing applied research inside new UMR's (\textit{Unité Mixtes de Recherche} / Hybrid Research Units) which are often administrated by the CNRS and designed to meet the needs of businesses.
\end{itemize}

\subsubsection{Private Research and Innovation Structures}
France has developed a lot in recent years its legal structures to support risk capital, like in the U.K where the same kinds of dispositions can be found. However, unlike in Germany, there are only in France a limited number of public-private partnership funds specialized in supporting small and medium enterprises (SME's) that foster innovating technologies : according to the \textit{Rapport de la Mission d'Évaluation Relative au Soutien à l'Économie Numérique et à l'Innovation} [1], only three such funds existed in 2012, but more are created continuously.\\

On the other hand, all the E.U states have created structures, whether through tax policy or legislative initiatives, to support research and development like Oséo in France, the Technology Strategy Board, Regional Growth Fund and Research Councils  in the U.K, the ZIM (\textit{Zentrales Innovationsprogramm Mittelstand} /Central Innovation Program) since 2008 in Germany and the Vinnova program in Sweden.\\

In France, a large selection of private research structures can be found and most have been created with the goal of unleashing the potential of public research and gear it toward innovation in the real-world economy and the private sector. Among those structures, we can give the following examples : 

\begin{itemize}
\item Private research centers that have been approved by the Higher Learning Education and Research Ministry and that can therefore receive public money.
\item \textit{Centres de Coordination de Recherche et de Développement} (CCRD's / Centers for the Coordination of Research and Economic Development) that have been created inside large multinational companies with the task of coordinating all research policies, planning research programs, administrating laboratories and of course maximizing all possible tax advantages for the company by optimizing all research paid for by the company with respect to existing government tax relief programs and research subsidies.
\item The status of \textit{Jeune Entreprise Innovante} (JEI / Young Innovating Enterprise) that can be awarded by the Finance Ministry to newly born SME's (even a one person company in the case of a start-up). Such a tax status confers many advantages to young companies on the condition that  they be eligible to the CIR (Crédit Impôt Recherche / Tax Relief for Research), about which we will talk more later and which is basically a tax deduction computed from the amount of money a company spends on its research activities, for at least 15 \% of their functioning costs. Moreover,  at least 50 \% of their capital must be held by private individuals, SCR's (\textit{Sociétes de Capital Risque} / Risk Capital Firms), FCP's (\textit{Fonds Communs de Placement / Mutual Investment Funds}), SFI's (\textit{Sociétés Financières d'Innovation} /  Financial Firms for Innovation), public research establishments or other JEI's. If those conditions are met, the JEI enjoys a complete exemption from corporate tax and social security contributions as well as a partial exoneration of capital gains tax for investors who will, at a given time horizon, sell their shares in the company.
\item The status of \textit{Jeune Entreprise Universitaire} (JEU / Young Enterprise with Academic Roots) which was created on the model of the JEI with the same tax relief advantages but which is specific to companies created by researchers working for a public research and higher learning institution. The JEU status has been designed to encourage professors, researchers  or even college students to become entrepreneurs and to monetize their research in order to derive extra income from it and create jobs in the real economy. A JEU has to be created by someone working in a higher leaning institution, has to have an economic activity based primarily on its creator's research and needs to be based on a convention signed between the higher learning institution and the new entrepreneur.
\end{itemize}

Despite all this, research expenses toward economic innovation in France remains limited in comparison with some other E.U states like Germany and the Scandinavian countries. This situation may not be however the consequence of a lack of dynamism in French research and development but merely an expression of the fact that the most innovating economic sectors in France (luxury goods, agribusiness and food processing, tourism, high value added services, etc..) aren't traditionally as much in need of scientific and technological research as the dominant economic sectors in countries like Germany.

\subsubsection{Business and Research Clusters ("\textit{Pôles de Compétitivité}")}
The business and research clusters, which are still too few in France compared to their numbers in the U.K and, especially, the United States are a key factor in the cooperation between the public and the private sector. Those \textit{Pôles de Compétitivité} consist of "areas", which can be physical or not, where private companies, higher learning establishments  and research institutions, both public and private, converge and pool resources to create a high degree of synergy for innovation, technological advancement and economic growth. In their respective fields of activity, those business and research clusters provide all the means of research as well as sources of financing so that new innovating companies can be born and existing businesses can thrive. In France, examples of such  \textit{Pôles de Compétitivité} can be found in Cap Digital for the IT sector and Finance Innovation for the financial sector. Their action is determinant in fostering innovation and economic growth in the country.\\

In France, the business and research clusters have been created since 2004 by the Comité Interministériel d'Aménagement et de Compétitivité des Territoires (CIACT / Inter-ministerial Committee for the Development and Competitiveness of the Regions) to support research programs approved by the State. Tax incentives toward the creation of such structures have been somewhat lowered since 2010 because their profits are no longer exempt from taxation but they remain exempted from all local taxes like the \textit{contribution économique territoriale } (regional tax to finance economic subsidies). Today, in most economic sectors from nanotechnology to finance, there are around 71 research and business clusters in France and 6 of them have a truly global perspective. These encouraging results remain however modest when compared with the business and research clusters, technological innovation centers and business incubators that can be found in the United States or, to a lesser extent in the U.K. As a matter of fact, those countries possess world-class giants like the Silicon Valley in California (6000 companies concentrated on 1.5 square kilometers and supported by Stanford University with its 15,000 students, including nearly 9000 post-graduates) or the East London Tech City (informally known as "Silicon Roundabout", it regroups 800 high-tech companies).

\subsection{A Flawed System, Comparison with other  Developed Countries}
Despite all these public and private structures and the high efficiency of some of them which confers to France a worldwide reputation of excellence in some fields of research and innovation, and despite a clear amelioration of the competitiveness in the past few decades when it comes to research and innovation in France, the system remains deeply flawed. Research and innovation in France remains burdened by many legal and administrative constraints and the \textit{Tableau de Bord de l'Union de l'Innovation} (Benchmark of Innovation in the Union) [3] which is published every year by the European Commission ranks France as 11th among the member countries, on par with the Netherlands, Belgium and the U.K as an  "innovation follower" rather than an "innovation leader" like Sweden (1st), Denmark (2nd), Finland (3rd) and Germany (4th). This disappointing position of France in the field of innovation, according to the European Commission, is however somewhat mitigated by its better results regarding the quality of its researchers and the number of its peer reviewed publications (7th and 8th, respectively, according to the same study). France has admittedly increased the performance of its research and innovation system between 2007 and 2014 but tends to evolve slower than its European partners because, according to the same study by the European Commission, it was 8 \% above E.U average in 2012 but only 6\% above E.U average in 2014.\\

This disappointing situation is confirmed by France's 14th position among E.U countries in the ranking of national economies by public-private partnerships. The insufficient amount of private sector investment in French SME's as well as the amount of risk capital available in the economy as a whole is also worrisome but this is an E.U-wide problem because, while the global amount of innovation in the European  economy has been stable over the past decades, the number of innovating private sector companies, especially SME's, is decreasing. The stable level of innovation at an E.U-wide level hides deep inequalities in the situation of individual countries. Indeed, the recent progresses of some states like Latvia, Bulgaria, Ireland or the U.K is compensating for the difficulties of others and therefore the E.U as a whole remains behind the countries that lead the world for innovation, like the United States, Japan or South Korea.\\

It has to be noted however that all those comparative studies are only taking into account the overall quantity of innovation in the economy and do not permit to draw any conclusions on the quality of the research and innovation structures of a country. Moreover, the profitability of the research and development  investments in France aren't properly measured either by those studies and to do so, one would need to use the theory of options and cash flows that those investments are capable of creating, as detailed in the work of Professor Raimbourg, published as part of the encyclopedic \textit{Ingénierie Financière} (Financial Engineering) book of reference published by Dalloz Action [4].\\

Another notorious flaw of the French research and innovation system is rooted in the French, and to some extend European as well, business culture. Indeed there is not enough of a “risk culture" in Europe : the business projects that get started must come to term otherwise it will be regarded as  a serious, even borderline shameful, failure. The bankruptcy and liquidation laws in France and Europe in general tend to be very punitive toward investors who didn't succeed and therefore,  any entrepreneur that starts a company that doesn't get off the ground is often considered to be a "loser" in European business culture. That situation  tends to diminish the dynamism of potential entrepreneurs who want to minimize the risks above everything else, even at the cost of missing business opportunities. This is totally different in the United States where it is much easier to rebound after a business failure. Excessive and overly complicated rules and regulations, as well as redundancy and sometimes even incoherence between individual member countries national regulations produces a lot of red tape that also, compounded with a very high cost of capital, tends to smother innovating businesses in Europe.

\section{The Financing of Innovation in Enterprises }
The financing of innovation in French and European enterprises is of three kinds : public financing by the State, private financing and European financing.

\subsection{Public Financing}
Public financing of innovation is essentially geared toward research. First of all, this public financing takes the form of tax reductions or exemptions. The \textit{Organismes Publics de Recherche} (OPR / Public Research Institutions) are entitled since the \textit{Loi de Programme pour la Recherche} (Research Planning Bill) of April 18th 2006 to a complete exemption of all taxation on any profits generated by their research activity within the constraints of their public interest mission. These provisions for the OPR concern primarily :
\begin{itemize}
\item The public research administrations like the CNRS and the public higher learning institutions like the public universities (for example, Paris I Panthéon Sorbonne University).
\item The legal entities created with the goal of administrating a research project or with the goal of coordinating the research activities of a network of other research institutions, including other OPR's. This is the case for example  of the \textit{Laboratoires d'Excellence} (LabEx / Laboratory of Excellence) that operate under the aegis of the ANR.
\item The research foundations that are recognized by the government as  \textit{Fondation Reconnue d'Utilité Publique} (Foundation Carrying out a Public Interest Mission) like for example the Pasteur Institute that is a world leader for medical research.
\end{itemize}
Moreover, the enterprises from the private sector are entitled to deduct the donations they make to the OPR's  from their taxable profits, within the limit of 0.5\% of their total revenue figure.\\

Besides those tax advantages, the traditional "vertical" structure of the French research and innovation system makes it so the public financing of research and innovation often takes the form of direct government subsidies or loans, subsidized or not, delivered to public, private or hybrid entities by a number of government agencies. Among the most important programs and institutions involved in financing innovation in France, we can cite for their important role :

\begin{itemize}
\item Direct subsidies provided, within the constraints of European Law which tends to limit such practices at the member state level, by the central French government through various ministries or by local governments at various levels (\textit{Municipal, Départemental, Régional},...)
\item Subsidies and loans to innovating companies as well as research grants offered by the large state-owned (or hybrid with a majority of the capital held by the State) companies like \textit{Electricité de France} (EDF / French Electric Power Company) or \textit{Société Nationale des Chemins de Fer Français} (SNCF / French National Railway Company).
\item The Agence de l'Environnement et de la Maîtrise de l'Énergie (ADEME /  Agency for the Supervision of Energy and Environmental Policy) which provides subsidies and loans to SME's and start-up companies that are innovating in the field of energy production and storage. It's main goal is to foster in France the development of green and renewable energy in the context of the energy transition currently taking place in all advanced economies world-wide and to make that transition profitable so that cleaner energy doesn't only make sense from an environmental point of view but from a business point of view as well.
\item State subsidies provided to firms that participate in the \textit{Conventions Industrielles de Formation par la Recherche} (CIFRE / Higher Education Through Industrial Research) program which consists of a company, from the large mutinational group to the start-up, providing a grant and one of his or her two research co-directors (the other research director has to be a university professor) to a doctoral student whose research is centered on the company's activity.
\item Oséo, which is legally a private bank specialized in providing loans to innovating SME's, but with a \textit{délégation de service public} (public service mission). It was merged in June 2013 with the \textit{Fond Stratégique d'Investissement }(FSA / Investment Planning Fund) and other State financial institutions  like the enterprise division of the  \textit{Caisse des Dépôts et Consignations} (CDC) to create the Banque Publique d'Investissement (BPI France / French Public Investment Bank).
 Oséo and the BPI are providing funding to innovating enterprises, especially young SME's through loans, loan guaranties and direct investments.
 \end{itemize}
All these state-controlled mechanisms to fund research and innovation in the private sector mustn't however hide the fact that it is the universities, public research administrations like the CNRS and other OPR's which are the key players in research and innovation in France. A similar  situation exists  in the United States, albeit with the major difference that most large universities, while often receiving some public funding, are essentially private institutions. In the United States, the law goes much further than in France and in most E.U countries in recognizing the key role of universities in research and innovation with the Bayh-Dole Act of 1980 that confers to universities the exclusive property of their research results. While French universities and OPR's do not usually enjoy such a level of independence from the State, they are still the keystone in the research networks that link the private sector, the public sector and the risk capital firms in order to convert scientific and technological advancement, obtained through research, into real economic growth and job creation.\\

Since the creation of the \textit{Programmes d'Investissement d'Avenir} (PIA / Programs for Future Investments) by legislative action on March 9th 2010, global State policy about innovation and research funding in France has been articulated around six major axis with the aim of creating a coherent research and innovation funding program that covers all aspects of scientific and technological progress including fundamental research, industrial development, as well as education and of course job creation and economic growth. Those six major axis are the following :

\begin{itemize}
\item Support higher learning and education with the goal of creating world-class research centers.
\item Foster the valorization of fundamental research through its applications in the economy with the goal of increasing and accelerating the transfers of ideas and people between private and public research and the creation of start-up companies by the researchers.
\item Provide financial incentives to the industrial sector so it can help support the development of innovating start-up's and SME's through training, know-how transfers and financial assistance.
\item Support the energy transition and the advent of clean and renewable energy production in France.
\item Support the digital economy and the development of a world-class telecommunication infrastructure in France.
\item Support medical research and the biotechnology sector.
\end{itemize}

The drafting and management of the PIA's is entrusted to the \textit{Commissariat Général à l'Investissement} (CGI / National Investment Commission) that coordinates all government programs in support of research and innovation in France and makes sure that all the concerned institutions (ANR, BPI, CDC, etc..) are acting in an efficient fashion to maximize the benefit to the economy and the public in general. The PIA's also play a key role in coordinating large scale scientific endeavors that require the pooling of the resources of many business and industry agents, research institutions as well as the support, both material and financial, of the State. Those large scale projects typically revolve around cloud computing, big data mining, a smart electrical grid, renewable energy, etc...\\

The advent of the PIA's has had a very strong positive influence on the French NSI and those programs help make up for other flaws in the French research and innovation system and helped France overcome many of the deficiencies of its traditional (and archaic) "vertical" system of research and innovation. The PIA's allow for an optimum management of private investment subsidized by public money and a better prioritization of the innovating business projects so that the State can better allocate its limited resources to the projects that have the best chances to translate research results into economic growth, the production of innovating products or services and the creation of jobs. The PIA's also permit a better coordination between academic and business agents and facilitates  exchanges between those two worlds.\\
 
The clear beneficial effects of the PIA's on the French NSI since the program was initiated tend however to fade nowadays according to the \textit{Cour des Comptes} (State Audit Office). Indeed, according to its December 2nd 2015 report [15], the first PIA (PIA1) which started in 2010 was faithful to the initial mission and did produce tangible results but the second PIA (PIA2) was launched in 2014 before the results of PIA1 could fully have come to fruition and before a proper macro-economic and statistical study of the effects of PIA1 could have been conducted. Moreover, PIA2 has sometimes been used to finance projects that may not have been relevant as innovating projects capable of producing economic growth. According to the \textit{Cour des Comptes}, up to 20\% of the investment spending under PIA2 may have been mismanaged. The current objective is to correct this course deviation by the advent of a third PIA (PIA3), which would have a better managed budget and tighter fiscal policy and which would be placed under the direct authority of the Prime Minister.\\
 
Besides considerations about the efficiency of the successive PIA's, their actual size in terms of investment by the State is also smaller than it seems and it tends to blur the budget evaluation of those programs. As a matter of fact the amounts of money really available to fund innovation in the economy (24 Billion euros for PIA1 and 10 billion euros for PIA2) are smaller than those announced (35 billion euros for PIA1 and 12 Billion euros for PIA2) because a sizable portion of the PIA's (around 9 Billion euros) take the form of non-expendable endowments, meaning that the public money is invested elsewhere and only the revenues generated by those investments are available to fund research and innovation. Taking all these facts into consideration, the adjusted global level of investment that the French government is providing to boost research and innovation in France has been stable at best since 2010 and it doesn't seem that the successive PIA's have measurably increased that amount. Moreover, despite the clear beneficial effect of the PIA's, according to a 2014 report by the OECD [2], France still lags behind other OECD countries in terms of public expenses for education, including higher education, expressed as a percentage of GDP ( France ranks 18th among OECD member countries) and in terms of public expenses for research and development (France ranks 19th among OECD member countries).

\subsection{Private Financing}
The form that private sector financing of innovation and research will take inside a given company and the means available to that company of obtaining funding, will depend essentially on its size. A difference has to be made between the funding of research and development by a company using its own funds or through investments coming from other private companies, and the risk capital companies or risk capital funds. With regard to the private funding of innovation, the French system includes most of the other mechanisms found in other advanced economies but their level of development and availability, especially regarding the availability of cheap capital for start-up's, often lacks behind what can be found in the NSI of world leaders of innovation like the United States. In the following section, we will explore the major means of private financing of research and innovation available to businesses in France, starting from the large listed companies whose shares are publicly traded on a stock exchange down the start-up.
\subsubsection{Companies that Have Access to the Stock Market}
Companies that are listed on a stock market are usually large corporate entities, which can also be conglomerates regrouping multiple large companies for tax reasons. They often have subsidiaries purely dedicated to research and development in their field of activity. Some of those subsidiaries may be based in France, especially those that specialize in fundamental research, but they are also often located abroad, again with the goal of lowering their taxes, in countries which have a more favorable patent and intellectual property regime than the French system. Those patents filled abroad may afterwards be sold or licensed, internally to their parent company, or to other companies operating in France. This raises the question of the transfer prices regarding the localization of profits generated by the selling or licensing of patents, which was studied by H.Hamaekers in a report for the Inter-American Center of Tax Administration [18]. For its part, France established a new committee, which is tasked with checking the fairness and lawfulness of those transfer prices. It is called the \textit{Mission d'Expertise Juridique et Économique Internationale} (MEJEI / Committee of Experts on International Legal and Economic Matters) and it was established by decree of the Finance Ministry on March 13th 2013. \\

Those large companies are the source of most of the research and development expenditures in the public sector. Indeed, according to a report by BPI France [16], only 96 large companies represent around 34\% of the research and development investment financed by the private sector in France. For those large companies, research and innovation is usually self financed using their own cash flows. Even though those large listed companies usually have extensive financial means at their disposal, the very high rate of the corporate tax (33,1/3 \%) in France tends to discourage investments in research and innovation which are often not regarded as absolutely essential to the business model of the company or to move those activities abroad where the rate of taxation may be lower. Indeed, as far as large listed companies are concerned, since they are funding their own research, considerations about the various regimes of taxation in France are determinant in their research and innovation strategy. In that regard, the most important points to consider are the following :

\begin{itemize}
\item \textbf{Tax regime for research toward patents}. Research and development costs incurred towards the creation of a patent or patented invention can be deducted from the taxable profit of a company, without any ceiling. This covers all the phases of production of the patent or patented invention, including fundamental research, technological development and product tests. This constitutes a very important financial advantage in terms of liquidity for a company as long as it conducts research with the goal of producing patents and new patented products.
\item \textbf{Tax regime for patents}. In France the tax regime for patents is relatively attractive when compared with other European countries like the U.K with its Patent Box system. The most important tax advantages in France regarding patents are the following :
\begin{itemize}
\item For patents that have been filled or acquired, it is permissible to deduct from the taxable profit (corporate tax or personal income tax depending on the legal structure of the company) an amortization equal to one fifth of the value of the patent each year during five years even though patents are protected by the \textit{Institut National de la Propriété Industrielle} (INPI / French Patent Office) for 20 years. This disposition is designed to increase the cash flow of businesses.
\item When a patent or patented invention is sold or licensed, a reduced taxation rate of 15\% is applied on the capital gain of the transaction instead of the usual 33,1/3 \% rate, except if the transaction is taking place between two companies which are related, for example a transaction between a subsidiary and its parent company. In case the sale price is lower than the actual value of the patent or patented invention, then the entirety of the loss can be deducted from the corporate tax or the income tax, depending on the legal structure of the business.
\item When a patent is added to the assets of a company in exchange for shares of its capital, then the company doesn't have to pay a registration fee even if the patent has already been exploited. 
\item For private individuals, the value of patents isn't added to the total wealth in the computation of the \textit{Impôt de Solidarité sur la Fortune} (ISF / Solidarity Tax on Wealth).
\end{itemize}

Despite this attractive tax regime for patents, the relatively heavy taxation of businesses in France (corporate tax, social security costs, etc...) has a clear negative impact on innovation. As a matter of fact, a common tax optimization scheme for a French company consists in conducting research in France in order to benefit from all the financial programs designed to lower its costs, then the patent produced by that research is filled abroad in ad-hoc legal structures based in a tax haven country or territory, finally that same patent is sold to the original French company which benefits from the tax deduction associated with the sale of a patent.
\item \textbf{The \textit{Crédit Impôt Recherche} }(CIR / Research Tax Credit). The CIR in France is one of the most generous tax deduction programs geared toward fostering research and innovation among all of the OCDE countries according to Daniel Boucher [6]. It is available for all companies and businesses in France, regardless of their size or legal form, as long as they have research and development expenses (fundamental research, applied research, experimental development and prototypes, etc...), whatever their field of activity. Since it is, as we have already seen, the large listed companies which account for most of the research and development expenses of the private sector in France, it is naturally those same listed companies that benefit the most from the CIR. The CIR is equivalent to the Research and Experimentation Tax Credit in the United States and the R\&D Tax Credit in the U.K but it doesn't have any real equivalent  in Germany that relies exclusively on other programs like the ZIM about which we have talked earlier and which is a very efficient tool to fund research and innovation in businesses, especially for SME's.\\

The CIR consists of a tax credit that concerns the taxable profit (corporate tax or personal income tax depending on the legal structure of the company). It represents each year one of the three largest losses of tax revenue for the State. If the tax credit is larger than the amount of the tax that is due, then the balance is reimbursed by State after at most three years or even sometimes immediately as, for example, in the case of a Jeune Entreprise Innovante (JEI). The CIR is equal to 30\% of the research expenses up to 100 million euros and 5\% above. Every year, the research expenses of a company are computed by summing the following items :
\begin{itemize}
\item The amortization of patents.
\item The salaries of researchers and the corresponding social security contributions.
\item A lump sum equal to 75\% of the amortization of patents plus 50\% of salaries of researchers and the corresponding social security contributions. This lump sum is designed to take into account all the other expenses related to the research activity in the company (rents, transports, computers, equipment, etc...).
\item The expenses corresponding to subcontracting research to another company or to a public institution, with a ceiling of 10 million euros.
\item Various other costs associated with research like INPI fees and technology watch.
\end{itemize}

On top of all these disposition, there is another tax credit called "CIR innovation" (Research Tax Credit for Innovation) that is reserved to SME's. It is a tax credit corresponding to 20\% of the expenses incurred towards ameliorating the efficiency of existing products without the intervention of researchers or the filling of new patents. The ceiling of the CIR Innovation is 400,000 euros per year.
 
 The CIR program is very expensive for the State. It costs around 5 billion euros per year in the form of lost tax income and this cost  is expected to grow up to 7 billion euros in lost tax income in coming years, when the program reaches its cruising speed. According to the OECD [2], this represents an amount four to six times larger than the combined government subsidies given to foster research and innovation in the private sector and one third of the costs incurred toward research and development in the public sector (which from a legal point of view do not include the CIR). Given its very large cost, the CIR program has naturally come under review and the question of its efficiency in supporting research and innovation in the French economy has naturally been posed many times in the political debate. While the CIR is undeniably an important program with a lot of potential to foster innovation in the French economy, and some impressive results over the years, as demonstrated by the work of the \textit{Observatoire du CIR} [19], it remains somewhat controversial. Indeed, the Cour des Comptes published a report in September 2013 [7] that underlines some flaws in the CIR system, and calls its  efficiency into question, especially given its high cost to the State. While also proposing many reforms to improve the efficiency of the CIR, some of the main conclusions of that report were :
 \begin{itemize}
 \item The efficiency of the CIR regarding its main objective, which is to increase the research and development investments of the private sector, is difficult to precisely quantify. Moreover, the increase in those research and development investments doesn't seem to reflect the tax advantage offered by the State.
 \item Managing the CIR is difficult, both for the companies that benefit from it and for the French tax administration. Much of the red tape could be removed by allowing online filling of CIR documents.
 \item The tax administration is unable to properly target its controls and tax audits in order to uncover companies which are abusing the CIR. A large part of the problem lies in the opacity and the complexity of the rules governing the CIR.
 \item The ministry of education and higher learning doesn't have the resources needed to assure in good conditions its mission of supervision of the research activities of the companies that benefit from the CIR.
\end{itemize}
\end{itemize}

\subsubsection{Companies that Aren't Listed on the Stock Market}
For those companies, which are typically of intermediate size, self funding of research and innovation is usually not an option because they do not possess enough own funds and the necessary cash flows generated by a high turnover. They aren't listed on a stock market which means that they cannot raise capital through the sale of shares on the stock market which remains the most efficient method of equity financing. They will encounter difficulties obtaining loans and guarantees to finance research and innovation from the banking sector as well because, especially in the current financial crisis climate, innovating SME's represent a very high risk for the banks. Because of the stringent banking regulations, at the French, European and international level with the Basel III recommendations, banks will find themselves in no position to offer loans to risky SME's. This fact is one of the main conclusions of a 2013 report by the European Commission [8]. Those financial and banking rules and regulations clearly serve an important purpose in defending the global financial and banking system against systemic events similar to those created by the Lehman Brothers collapse of 2008, which was precipitated by the Sub-prime Crisis taking its roots in the U.S real-estate sector  where banks and other financial institutions took the unfortunate habit of offering a very large number of bad and extremely risky loans to struggling people who had little chance of ever repaying them. However, effectively forbidding banks to offer risky loans stifles innovation in the economy as well. Leverage ratios and own funds requirements of Basel III in particular tend to have the pernicious effect of starving innovating SME's of the much needed capital they need to develop their activity and/or of rendering access to capital too expensive for them, because the cost for the banks of providing that capital has become to high. A middle ground has to be found in the future between security and avoiding smothering innovation in the economy because while all those new banking regulations might help prevent another Lehman Brothers, they might also unfortunately kill in the egg the next Google or Facebook.\\

To remedy this situation, France has created many legal and financial structures designed to provide much needed capital to innovating companies of intermediate size so that they can develop their activity. Those many private equity initiatives and venture capital trusts are sometimes more efficient that stock markets in providing capital to innovating SME's created by dynamic entrepreneurs, according to R.S Harris, T. Jenkinson and S.N Kaplan [9]. Among those structures, the most notable are : 

\begin{enumerate}
\item The \textit{\textbf{Fonds Communs de Placement pour l'Innovation}} (FCPI / Mutual Investment Funds for Innovation). Those structures are collective investment funds that are neither companies nor trusts from a legal point of view. They are basically just an "account" that is being managed by a bank or another financial institution and they are a particular kind of Fond Commun de Placement (FCP). The structure of an FCP and the steps toward its creation are always the same, be it designed for innovation in the economy or not :
\begin{itemize}
\item First, a banking establishment creates an account (the FCP) which is designed to receive deposits from both private investors and businesses. The legal form of the FCP specifies the ways the bank will have to manage it. For example, the bank will have to  invest the deposits into a given sector of activity (innovating companies for the FCPI), or into the economy of a given country or territory, or into a given kind of businesses, etc.. 
\item Investors then place their money in the FCP and receive in exchange shares of the FCP, which are expected to increase in value with time and which may be sold later on, generating a capital gain.
\end{itemize}
At least 70\% of the assets of an FCPI must be invested in the capital of unlisted companies of less than 500 employees and which are considered as being "innovative" according to the following criteria :
\begin{itemize}
\item They must have research expenses eligible for the CIR representing at least one third of their total revenue.
\item They have to be able to provide evidence for the creation of innovating products, techniques or services.
\item At least 70\% of their capital must be held by private individuals, or a public research establishment, or other innovating companies of the same kind, according to the same criteria that we have just detailed.
\end{itemize}

In order to support the development of the FCPI's, a special tax reduction system has been put in place for private individuals who invest in them by purchasing shares. The main provisions of this tax reduction program are the following :

\begin{itemize}
\item  There is first of all a tax credit (informally called "\textit{réduction Madelin}" in the name of a former French finance minister of the mid-1990's, \textit{Code Général des Impôts} (CGI) /Tax Code, art.199 terdecies-0A), that is of the same nature as the British Enterprise Investment Scheme (EIS), and which is equal to 18 \% of the yearly investment into FCPI shares, with a ceiling of 50,000 euros per year for a single person and 100,000 euros per year for a married couple, under the condition that the shares be kept for at least five years.
\item Then there is a tax advantage known as "\textit{non-imposition immédiate}" (delayed taxation) which consists of adding the revenues and capital gains generated by the FCPI shares to the value of those shares (which may be sold at a later date) without immediate taxation by way of the personal income tax.
\item Finally, if the shares are kept for at least five years, there is a complete exemption of the personal income tax corresponding to capital gain generated by the sale or acquisition of those shares.
\end{itemize}

Moreover, these FCPI structures allow investors to benefit from the experience of professionals specialized in funding research and innovation in the economy and in handling innovating companies as well as the inherent risks associated with them. F. Moulin and D. Schmidt detail in their book "\textit{Les Fonds de Capital Investissement, Principes Juridiques et Fiscaux}" [10] how investing in FCPI's permits to benefit from people with knowledge of the innovating SME's business ecosystem, to achieve economies of scale and to pool the risks.\\

Despite the many tax advantages, the FCPI structure seems however to be losing its attractiveness among investors. Indeed, there were around 97,000 owners of FCPI shares in France in 2014, down from 145,000 in 2008 and the total amount invested in FCPI's has dropped by 32\% during the same period according to A. Vion in Les Echos newspaper of April 16th 2015 [11].

\item \textbf{Other private equity structures}. Besides the FCPI's, many other structures, which may differ among one another with regards to their size, scope and target companies, exist in France to finance the research and innovation of the SME's. They all revolve around private equity and can take various legal forms. All these legal structures benefit from a tax exemption on their profits (exemption of the corporate tax or of the personal income tax depending on their legal form).
\begin{itemize}
\item The \textbf{\textit{Fonds Communs de Placement à Risque}} (FCPR / Mutual Fund for Risky Investment) Those funds have essentially the same legal form as FCPI's but they are specialized in "risky investments", which in particular can be investments in innovating companies.
\item The \textbf{\textit{Sociétés de Capital Risque} }(SCR / Risk Capital Firm) Those firms are financial companies specialized in asset and portfolio management. Their mission is to help the launch of new companies which are active in research and innovation by investing at least 50 \% of their assets in the capital of unlisted innovating SME's or in the capital of innovating companies with a small capitalization and limited access to the stock market.
\item The \textbf{\textit{Sociétés Unipersonnelles d'Investissement à Risque} } (SUIR / One Person Company Dealing with Risky Investment) Those financial firms are similar in structure to the SCR's but they consist of only one shareholder who has to be a private individual.
\item The \textbf{\textit{Sociétés Financières d'Innovation} }(SFI / Financial Firm for Innovation) Those firms, which have to be registered with the Finance Ministry and are subjected to government oversight, support research, innovation and technological advancement in the economy by funding the development of SME's involved in the elaboration of new products and services. They must invest at least one third of their assets in innovating projects. Their investment can take the form of a direct acquisition of shares of the capital of innovating SME's or they can provide to them intellectual property assets like patents relevant to their innovation.
\end{itemize}
\end{enumerate}

Moreover, besides the "\textit{réduction Madelin}" that we mentioned earlier, there is also for private individuals a reduction of the \textit{Impôt de Solidarité sur la Fortune} (ISF / Solidarity Tax on Wealth) equal to 50 \% (CGI, art. 885-0 bis) of the investment in shares of the capital of SME's from an E.U country (not only innovating SME's) that operates in the industrial, commercial, arts and crafts, agribusiness or service sectors, but not in the financial sector. The ceiling for the ISF reduction is of 45,000 euros per year and this tax deduction cannot be cumulated with the "\textit{réduction Madelin}" on the personal income tax.\\

Despite all these programs and structures designed to boost private investment into innovating SME's, the French system is still far behind the American system in terms of efficiency and dynamism. Indeed, France doesn't have in its legal system something as versatile as the status of Small Business Investment Company (SBIC). SBIC's are basically private businesses specialized in investing in SME's and which can use their own funds as well as loans that have been guaranteed by the federal SBA (Small Business Administration). In the American business landscape, one can also find the New Market Venture Capital Companies (NMVCC) which specialize in investing in the capital of SME's located in economically underprivileged regions of the U.S, and which also support research and innovation. In the U.K, there used to be the Capital for Enterprise Limited (CfEL)  until it was merged in October 2013 with the  Department for Business, Innovation and Skills (BIS) to be part of the British Business Bank program. It continues under this new form its mission of providing financial assistance to British SME's and start-up companies and the program is considered to be close to being as successful as its American counterparts. In France, progress still needs to be made and the Finance Ministry underlined in its Report of the Committee for the Development of the Economy and Innovation in the Digital Age [1] that :
\begin{itemize}
\item When it comes to funding research and innovation in the economy, there is still in France an insufficient role played by the "business angels" and venture capital (France is even below the average of the European Union for the availability of venture capital expressed as a percentage of GDP) and that situation persists despite the creation of the \textit{Fond National d'Amorçage} (FNA / National Fund for Start-up Companies) which represents only around 100 million euros of investment per year. This may have to do with legal restrictions (for example the obligation in some cases to sell the start-up to a European company) that may apply when the entrepreneur attempts to sell a successful start-up that benefited in one way or another from public funding. Indeed it is often the business strategy of start-up company creators to intend from the beginning to sell the company later on to an industry leader (to give a few very famous examples: Facebook bought What's App, Google bought Youtube, etc...) and this sale is often the only real compensation the original entrepreneurs gets for their work because in the growth phase, they often get no salary from the fledgling company they created.
\item The availability of risk capital is still insufficient in the "middle stage" of the life-cycle of a typical start-up company. That is the investment that is supposed to take over the funding of the start-up after an initial phase of funding, led by the original investors themselves and their families.
\item Technical and legal hurdles need to be overcome to permit successful start-up companies at a later development stage to access the stock market more easily, thus raising the prospect of raising more capital or being acquired by a larger firm in France.
\end{itemize}

\subsubsection{The Case of the Start-up Companies}
Start-up companies are usually very small companies (\textit{Très Petites Entreprises} (TPE)) which are born around "key-persons" (sometimes a single entrepreneur) who usually plan to exploit the economic potential of their own research, patents or patented inventions. The typical economic and financial strategy behind the setup of a start-up company based on original scientific research is the following:
\begin{itemize}
\item Researchers or inventors provide their patents, patented inventions or know-how to the newly created company that starts to develop around those "key persons". There is something like a bet in a project like this : if the company succeeds and grows, then the key persons will get rich by selling their shares of the capital later on, but if the company fails the loss will be theirs and they will get nothing.
\item The compensation of the key persons usually doesn't take the form of a salary which is much too expensive from the point of view of a small company because of all the taxes and social security contributions associated, and not attractive enough from the point of view of the entrepreneur. This compensation will, hopefully, take the form of the capital gain generated from the sale by the key persons of their participation in the company (their shares of the capital). Once the key persons have severed their link with the company they created, they sometimes start over and create another company and attempt to be, in the jargon of the start-up creators community, "serial winners".
\end{itemize}
In order to facilitate start-up creation, which is associated with innovation and dynamism in the economy, France created a series of legal protections, financial structures and instruments as well as a special legal status for start-up creators, much like the United States with the Employee Stock Ownership Plans (ESOP). The key points of the French system regarding start-up creation are the following :
 \begin{enumerate}
 \item Upon entry of the inventor/entrepreneur into the new start-up (or into an existing start-up that wishes to develop a new product using his or her research), he or she brings a patent, patented invention or know-how and may receive in exchange several kinds of financial instruments like :
 \begin{itemize}
 \item The \textit{Obligations Convertibles en Actions} (OCA / bonds convertible into shares) or the \textit{Obligations Échangeables en Actions} (OAE / bonds exchangeable for shares). The precise workings of those financial instruments are explained in details in the work of Jean-Claude Augros [12] and of Philippe Raimbourg [5], but the basic idea is for the inventor to enter the company as a mere lender and to exit as a shareholder.
 \item \textit{The Bons de Souscription de Parts de Créateur d'Entreprise} (BSPCE / share warrants for entrepreneurs) which enable the future creation of new shares earmarked for the key persons of the start-up company.
 \item The \textit{Attributions Gratuites d'Actions} (AGA / attribution of free shares) that are reserved for the key persons upon creation of the start-up company. Their precise workings are detailed in the work of P. de Fréminet [13].
 \item The \textit{stock-options} (SO) that are designed to provide compensation for the key persons by giving them the right, upon creation of the start-up, to buy shares of the company in the future at a given strike price (call option). Assuming the value of the shares will increase over the years, the SO allows the key persons to realize a future capital gain. Obviously, the risk lies in a depreciation or stagnation of the value of the shares.
 \end{itemize}

All these dispositions therefore permit to compensate an inventor/entrepreneur, not with a salary but with a future capital gain on the sale of shares, when he or she eventually leaves the start-up. This capital gain corresponds to the added value the inventor/entrepreneur provided to the start-up through his or her research, innovation and management skills. It has to be noted that this compensation arrangement has been for a few years less and less profitable because it is based on the difference between the taxation levels of capital gains and salaries. Indeed, salaries tend to be more expensive from a tax optimization point of view and they also include social security contributions. However, this difference between the levels of taxation of capital gains and salaries are currently being called into question and many of the tax advantages of the usual compensation system of start-up creators are in the process of being adjusted downwards, as the levels of taxation of capital gain and salaries converge in many cases. This analysis is detailed in the work of A.Guillemonat, O.Ramond who published an in depth review of current management packages in the 10th issue of the \textit{Revue de Droit Fiscal} in 2015 [14].

 \item Upon exit of the inventor/entrepreneur from the start-up company, the sale of his or her shares generates a capital gain. This capital gain is admittedly subject to taxation under the same terms as dividends from equity (progressive rate of 0\% to 45\%, plus social security contributions), except for BSPCE which are taxed at a flat rate of 19\% plus social security contributions, however a tax deduction is offered depending on the duration the shares were kept by the  inventor/entrepreneur. This tax reduction can climb up to 85\% for shares of an SME which have been acquired within ten years of the creation of the company. Moreover, upon exiting the start-up and selling his or her shares, the  inventor/entrepreneur may claim further tax reductions under French law by invoking the taxation regime that is reserved to the "\textit{revenus exceptionnels}" (exceptional revenue). Today, in order to benefit from those dispositions, the key persons aren't required to have top management responsibilities inside the star-up or to be officially heading the company, as it was mandatory in the past.
 \end{enumerate}
 
Many other structures, which may be public, private or hybrid support the creation of start-up companies, in every field of activity, in France and the innovation and dynamism they bring into the economy. To cite only a few, we can talk about :

\begin{itemize}
\item The important role played by the \textit{pôles de compétitivité} and the \textit{pôles d'innovation} (business and research clusters) as we have already seen in the part about the tools of innovation and economic growth. Those business and research clusters provide funding to the fledgling companies but their action isn't limited to that and they do as well offer legal advice, practical and technical advice and support for the entrepreneurs or would-be entrepreneurs who have the project of exploiting their innovating ideas by creating a start-up. In France, Cap Digital fulfills this mission in the information technology sector and Finance Innovation fulfills this mission for start-up companies in the financial sector, especially supporting the newly born companies through its program of "\textit{projets labélisés}" (approved projects) which gives financial start-up's the recognition and visibility they need to thrive in a very competitive business landscape.
\item The emerging role played by the "crowd funding" companies and online platforms. As a matter of fact, despite the many funding opportunities that we have seen, sometimes an entrepreneur cannot have access to private equity, government subsidies or loans and can only count on his own funds to start a company. This is where the crowd funding solutions available in France may play a determinant role. These crowd funding solutions (Anaxago, Wiseed, Finance Utile, etc.) serve as intermediaries between "micro-investors" and the would be founder of a TPE. They are like the middle ground between a proper investment fund and family investment. They have also the advantage of relying on professionals with knowledge of the TPE ecosystem, like the \textit{Conseillers en Investissement Participatif} (crowd funding advisors), which is a label recognized  by the Finance Ministry, who can vet the projects and provide valuable advice to both the investors and the entrepreneurs.
\end{itemize}

\subsection{European Financing}

\subsubsection{European Programs}
The funding of the European research and innovation policy has been since 1984 articulated around the Programmes Cadres de Recherche et Développement Technologique (PCRDT / Framework Program of Research and Technological Development), but the European research policy only reached its true potential with the creation in January 2000 of the Espace Européen de la Recherche (EER / European Rsearch Area) by the European Commission under the initiative of Philippe Busquin, who was then the European Commissioner for research and who authored several reports underlining the deep flaws in the European research system and its inability to transform scientific research into economic growth. Over the years, the greater role taken by the European Investment Bank (EIB) in funding research and innovation also accounts for the much more efficient nature of the research and innovation system in the European economy today.\\

The 8th PCRDT named "\textit{Programme Cadre pour la Recherche et l'Innovation}" (Framework Program for Research and Innovation) is part of the \textit{Europe 2020} program, which is a 10-year strategy proposed by the European Commission on March 3rd 2010, and it covers the period 2014-2020. According to an official report authored by José Manuel Barroso for the European Commission [17], the goals of the program are to maximize research and innovation funded by the European Union, to foster economic growth that is sustainable, inclusive and employment-generating and to face the great challenges in the European society and economy. Toward those goals, the program must create a coherent series of structures and institutions spanning the entire research and innovation ecosystem from fundamental research to the introduction on the market of new innovating products and services.  \\

The objectives and the budget of each PCRDT are set by the Conseil Européen de la Recherche (ERC / European Research Council) under the direct supervision of the European Commission, which calls upon independent experts for the selection phase and the evaluation of projects, which are  chosen after a public invitation to tender and are published afterwards in the Official Journal of the European Union. The 7th PCRDT lasted seven years had a budget of 50.5 billion euros, to which should be added 2,7 euros for the legally separate, but structurally integrated, Euratom project on nuclear research between 2007 and 2012. The 8th PCRDT is envisioned to have a budget of 80 billion euros until 2020, including estimations of 25 billion euros for cutting edge scientific projects, 18 billion euros for industrial innovation which includes nanotechnology, nanoelectronics, etc., and 32 billion euros for major challenges in European Society which include public health, agriculture, clean energy and the energy transition, marine research, etc.

\subsubsection{The European Investment Bank}

The \textit{European Investment Bank} (EIB) has been established in 1958 under the Treaty of Rome. It is a non-profit international financial institution whose shareholders are the E.U member states. It functions as a long term lending institution and its primary mission is to financially support economically sound projects that are important for European integration and the future of the European Union. The EIB is the main shareholder of the \textit{European Investment Fund}  (EIF), possessing 62\% of its capital. The EIB is taking a very active role in support of European Commission policies, especially the current \textit{European Commission Investment Plan for Europe} (ECIPE), known informally as "Junker's plan" and in that framework it crated the \textit{European Fund for Strategic Investment} (EFSI).\\

The role of the EIB and of its satellite organizations (EIF, EFSI, etc..) in the funding of innovation in Europe is crucial and represents a large part of the bank's activity even though it has other important missions, outside the scope of this study, like financing infrastructure projects in Europe (bridges, hospitals, etc...) and facilitating European integration and economic stability in Europe. The role of the EIB in fostering innovation in Europe became even more important in the context of the current global financial crisis. Indeed, the funding of innovation has suffered a lot after the 2008 financial crisis, that started in the summer of 2007 as the Sub-prime crisis in the United States. In the first semester of 2009, for example, there was almost no credit offered by the banks, which brought the real economy and innovation in particular to a standstill. The E.U member  states intervened to prevent the banks going under by establishing programs similar those of the American Federal Government (quantitative easing, etc..) that were deeply unpopular because they were seen by the general  public as a gift made to the banks while in reality it wasn't: those were loans offered to the banks from the states and the banks did pay them back in full in the years that followed, which even generated a profit for the states. In fact, since those loans were a lot more expensive than inter-bank lending, had it been available at the time, banks were in a hurry to reimburse the states and move on. Even though those rescue packages for the bank were successful in preventing a total banking and financial collapse, the availability of credit has never fully recovered to this date in France and in Europe, and innovation suffers a lot because of this.\\

In this context of a credit crunch, it is the SME’s that suffer the most because they are the ones that would be quickly asphyxiated in the absence of funding. One of the principal missions of the EIB is to help those SME’s find funding; not by directly giving them loans as it doesn't have the infrastructure and many local branches needed to interact with individual clients directly, but by providing traditional commercial banks incentives to lend to those SME’s, especially innovating SME's. The EIB can provide credit to banks on the condition that they use those guarantees exclusively to lend to SME’s. Those SME's need of course to have a sound business project and they have to be able to reimburse their loan since the EIB, while it's a not for profit organization, isn't an institution that hands out economic and financial aid either. Those EIB guaranties make lending to innovating SME's and start-up's an activity that is more attractive, less risky and more profitable for banks. The guaranties induce several beneficial effects for banks and they also generates leverage, which is typically quite high and in the order of a factor of 8 to 15, but remains safe because it is powered not by mere speculation but by the status of the EIB as a trustworthy European Institution, on the same level of standing and trust as the European Central Bank (ECB).  \\

Specifically, the advantages that banks get from EIB guaranties in exchange from offering loans to SME's, especially innovating ones, can be listed as follows : 

\begin{itemize}
\item A \textbf{rate advantage}:  because lending to young unproven companies, especially those that are producing innovating products, which may on may not succeed commercially, is more risky for a bank, it may refuse to do so in general, therefore young innovating companies do not get the credit they desperately need to grow and succeed. The EIB guarantee (usually 40\% to 50\% of the value of the loan) allows the bank to be compensated for the risk it is taking and to finance that risk. Because of that guarantee of the EIB, the bank can lower its lending rate or take more risk at a given rate, which is currently still quite low and fixed by the ECB.
\item An \textbf{equity advantage}: offering credit to sub-prime agents has become considerably more expensive for banks since Basel III regulations started to come into effect, accompanied by a great increase in the complexity of the rules and regulations for the banking sector, which are becoming more and numerous every year. There have been for example 47 separate new rules in France in 2009 for the banking sector alone. Because those prudential regulations force banks to have ever increasing levels of equity to back up their loans, which is crippling and expensive, banks will greatly prefer to lend to SME’s that have a low risk of failing (a baker's shop for example) rather than to the innovating ones, created by an entrepreneur who believes he or she has a revolutionary product based on new research. Even if the entrepreneur who believes he or she has a revolutionary product could in theory be the founder of the new Google or Facebook, banks just cannot take the risk of lending money to that person :  in the current regulatory atmosphere it just wouldn't make business sense to them. The guarantee from the EIB allows banks to lower their risk to the level where it makes good business sense to them to use their precious equity, demanded by the regulator, to back up a loan to an innovating company, thus fostering research and innovation in the economy.
\item A \textbf{balance of payment advantage}: this is directly linked to the regulatory requirement for the banks of having equity to back up the loans. Indeed the loans that are guaranteed by the EIB and which are, in the bank's view, potentially toxic because they are offered to small unproven, often innovating companies, can through legal mechanisms be, at least partially, removed from the bank's balance of payment, thus eliminating the need for the bank to commit more equity and own funds to back them up. In the context of the current financial crisis, large banks usually do not have liquidity problems, it is the regulatory high level of equity needed to back up loans which is the limiting factor to providing credit to the real economy and especially to the innovating agents of that economy. Therefore the banks need the guaranties offered by the EIB much more than they would need extra liquidity, provided by the EIB or the European Central Bank (BCE).
\end{itemize}

The EIB can also help fostering innovation in Europe through its participation in the EIF. Unlike the EIB which specializes in very large, multi-billion euros operations with commercial banks and doesn't deal with the SME’s in a direct fashion, the EIF has a more local approach and specializes in smaller operations focused on individual business projects which go though a rigorous vetting process. In most cases, the EIF will act as a fund of funds: taking a participation in specialized risk capital funds that in turn provide financing for very innovating, cutting edge start up companies. In that respect, the EIF is acting a lot like a business angel or venture capitalist, except that its primary motivation as a European Institution is to make those companies succeed for the greater good of the economy, rather than short or medium term profit and in that regard, the EIF also forfeits most of the rights that a traditional investor would demand, like the right to have a say in the business decisions and general strategy of the companies it invests into. The specialized funds that the EIF is working with have the know-how and the experience to properly evaluate the chances of success and the soundness of the business plan of those start-up companies in their field of expertise. It is a very hands-on, very local business to vet those projects which may be very cutting edge and based on a newly filled patent. These specialized risk capital funds (around 45 of them in France and 460 in the European Union) provide much needed capital investment to innovating start-ups (maybe the next Google), which are too small to have access to the stock market and which are also too risky to have access to bank loans.\\

Finally, the EIB finances innovation in Europe in the framework of the Junker's Plan through the EFSI, which is designed to work in concert with European Commission economic policies. For example the EIB has intensified its action, especially toward SME’s and innovating SME’s in the framework of the Junker Plan. As a matter of fact, in 2008 it issued  57 billion euros of guaranteed loans to banks among which 30\% were earmarked for the SME’s and in 2014 that commitment had increased to 77 billion euros of guaranteed loans to banks, again, 30\% of which were earmarked for the SME’s. The EIB and EIF also launched together the InnovFin initiative in the framework of the European Commission's \textit{Horizon 2020} program, which is an E.U wide research and innovation program with nearly €80 billion in funding between 2014 and 2020, as we have already seen. InnovFin consists of financing tools as well as advisory services designed to boost research and innovation by attracting investors and thus help all companies, from the start-up to the multinational, that wish to participate in the program, to be more innovating. The EIB also helps supervise and finance many \textit{Public Private Partnership} (PPP) projects, which are an important vector of innovation in the European economy.

\section{Conclusion}
As a conclusion, we can say first of all that, when it comes to research and innovation that foster sustainable economic growth and of the creation of jobs in the economy, both France and the European Union have progressed a lot in recent years. While some worrisome flaws remain in the French NSI, structural reforms in the framework of European integration, and often on the American model, keep on improving the situation. While France may not be ranked by the European Commission as one of the top innovating countries in Europe at the moment, as we have seen, France does remain a key player with world-class research institutions and dynamic entrepreneurs. To come back to the six modules that constitute an NSI, we could say that France has excellent individual parts but that the interactions between those parts, sometimes, still need to be perfect today. Regarding \textit{\textbf{human resources}}, France still lags behind other European countries when it comes to offering young people an education that will enable them to be competitive in an innovating economy and to give them the opportunity to be innovators themselves. Of course, the French higher education system produces world-class researchers and engineers, in all fields of science, and the French business schools are ranked among the best in the E.U and the world, but this excellence "at the top" is not enough. A healthy innovating economy also needs qualified technicians, accountants, marketing and communication people, etc.., whose work is essential for research and innovation to be translated into successful companies, from the start-up to the multinational, and economic growth. This is the kind of people that the French education system, despite recent reforms in the good direction, seems incapable to produce, despite high unemployment in the country.\\

Regarding \textit{\textbf{public research}}, France with its many large and world-renowned public research administrations, like the CNRS, is at the cutting edge of science and technology in many fields, but there is still, despite recent progress in the right direction, an insufficient mobility, both of ideas and of people, between the world of public research and the world of private research, business and innovation in the economy. \textit{\textbf{Private research}} is in good shape as well, as an NSI module of its own, and it benefits from a lot of help from the French State and the European Union as well. However, the very favorable tax regimes and various financial structures designed to foster research and innovation in France and in Europe cannot fully compensate for the very heavy taxation and social security contributions that are burdening businesses, stunting investment and stifling innovation in the economy at the French and European level. Of course it is not that simple but, taking a few shortcuts, we see very often a situation like this: scientific research is conducted in France while corresponding economic innovation is taking place in countries with lower taxes. Steps are taken, both at the French and European level to counter situations like this, but more needs to be done. Moreover, as we have seen, the majority of private research and innovation is still done by a handful of large companies, which are precisely those having the means to fully exploit every French tax reduction program like the CIR, while at the same time having the means to move production abroad where taxes are lower and/or labor legislation laxer than in France. Innovating SME's and start-up companies are having a more and more important role in private research, supported by business and research clusters like Finance Innovation or Cap Digital, French institutions like BPI France and European Institutions like the EIB, but despite remarkable progress, especially given the current global financial crisis, a lot still needs to be done to help innovating SME's and start-up companies thrive.\\

When it comes to the \textit{\textbf{relationship between industry and science}}, the French system still retains to this day many characteristics of its traditional, and obsolete, "vertical" structure. That traditional French system was based on research and scientific policy being decided at the top political level and articulated around traditionally State controlled industries like national defense, nuclear power, aeronautics and space, etc... Today, the relationship between industry and science is becoming more "horizontal" in France, and research is becoming better adapted to the needs of the economy, in every fields of economic activity, especially as industrial research and development partnerships are created involving companies, large and small,  from other countries inside the European Union.\\

\textbf{\textit{Innovating Entrepreneurs}} are many in France and their dynamism is world-renowned and an important engine of economic growth. As we have seen, many highly efficient structures and programs exist in France and at the European level that help the creation and continued funding of start-up companies as well as small and medium enterprises. There are attractive tax reductions systems for start-up creators, many private equity structures that also benefit from tax advantages, the business and research clusters, which offer not only financial help but legal and material help also, and the action of the European Investment Bank (EIB), which enables commercial banks to offer loans to risky SME's and start-up's in a very difficult global financial situation. Despite all this, a lot still needs to be done in France to facilitate the work of innovating entrepreneurs and to render the French system as attractive as the American or British system regarding the creation of start-up companies: in France, a lot of legal and administrative roadblocks on the path of innovating entrepreneurs still remain, unfortunately.\\

Finally, \textbf{\textit{general State policy}}, the last NSI module, is supposed to be cement that binds together all the other parts into a coherent whole. Many countries have had a highly efficient global State strategy for research and economic innovation for decades. This is the case in United States, with the \textit{Office of Science and Technology Policy}, which advises the President and helps define general federal policy regarding research and innovation. This is also the case in Germany and the U.K. France used to lag far behind in that regard and for decades, general State policy for research and innovation in France remained either essentially absent or tainted with short-term political maneuvering that completely lacked vision and a long term sustainable plan to boost research and innovation in the French economy. This situation has however been greatly improving recently, especially with advent of the \textit{Programmes d'Investissement d'Avenir} (PIA's), in concert with the process of European Integration. At the European level, France is naturally part of the \textit{Europe 2020} program, started in 2010, that fixed precise objectives, for individual member countries and the E.U as a whole, regarding research and innovation in the economy. There is for example the goal of reaching a figure of 3\% of the GDP for research and development investments, while it stands at 2,5\% in France today. There is also the goal of having at least 40\% of new generations graduating from a higher leaning institution and obtaining a diploma. This "union of the innovation" is under the supervision of the European Commission, which may propose objectives and make yearly recommendations to the member States regarding research and innovation. It is inside this European framework and through partnerships with other E.U countries that the French general State policy has the best chances of finally becoming the cement binding together all the remarkable assets that France possesses in its NSI. Indeed, it is as a part of Europe, that France will eventually achieve its true potential as a world leader in scientific research and economic innovation.

\end{document}